\documentstyle[twocolumn,prl,aps,psfig]{revtex}
\begin{document}
\preprint{}
\draft
%
%%%%%%%%%%%%%%%%%%%%%%%%%%%%%%%%% TITLE PAGE
%
\title{Optimal Monitoring of Position\\
in Nonlinear Quantum Systems}
\author{Michael B. Mensky}
\address{P. N. Lebedev Physical Institute,\\
Russian Academy of Sciences, Moscow, Russia 117924}
\author{Roberto Onofrio}
\address{Dipartimento di Fisica ``G. Galilei'', Universit\`a di Padova,\\
Via Marzolo 8, Padova, Italy 35131}
\author{Carlo Presilla}
\address{Dipartimento di Fisica, Universit\`a di Roma ``La Sapienza'',\\
Piazzale A. Moro 2, Roma, Italy 00185}
\date{Phys. Rev. Lett. 70 (1993) 2825-2828}
\maketitle
%
%%%%%%%%%%%%%%%%%%%%%%%%%%%%%%%%% ABSTRACT
%
\begin{abstract}
We discuss a model of repeated measurements of position in a quantum 
system which is monitored for a finite amount of time with a finite 
instrumental error.
In this framework we recover the optimum monitoring of a harmonic
oscillator proposed in the case of an instantaneous collapse of the
wavefunction into an infinite-accuracy measurement result.
We also establish numerically the existence of an optimal measurement 
strategy in the case of a nonlinear system. 
This optimal strategy is completely defined by the  
spectral properties of the nonlinear system.
\end{abstract}
%
%%%%%%%%%%%%%%%%%%%%%%%%%%%%%%%%% PACS NUMBERS
%
\pacs{03.65.-w, 06.30.-K}
%
%%%%%%%%%%%%%%%%%%%%%%%%%%%%%%%%% PAPER CONTENT
%
\narrowtext
Improvement in the precision of measurements  brings to consider the ultimate 
limits of sensitivity imposed by quantum mechanics and to 
develop measurement strategies overcoming such limits \cite{BRAG}.
A firstly proposed example of these strategies, 
also called Quantum Non-Demolition (QND) measurements, was the
stroboscopic measurement of position in a harmonic oscillator. 
A series of ideal infinite precision and instantaneous
measurements performed each half period of a harmonic oscillator
represents an optimal measurement strategy with perfectly predictable
results \cite{BRAG1,BRAG2,THOR}. 
In a realistic scenario it is compulsory to study a strategy based on
measurements which are affected by an instrumental error and which last a
finite amount of time. 
Besides this generalization, as outlined in \cite{BRAG}, 
quantum measurement models for nonlinear systems, {\it i.e.} systems 
which are not a harmonic oscillator, are still missing. 
In this Letter we study optimal strategies for measuring position in
nonlinear systems monitored for a finite time with finite accuracy.
By using the path-integral approach 
to quantum measurements \cite{ME,GA} we quantitatively recover the 
results for the QND stroboscopic measurements of a harmonic oscillator and 
we establish the existence of an optimal monitoring for a nonlinear system. 

The standard quantum limit in a continuous measurement
of position for nonlinear systems has been already analyzed 
in the framework of the path-integral approach \cite{MEOP}. 
The measuring system is schematized by an arbitrary 
measurement output $a(t)$ and an instrumental error $\Delta a$.  
The effect of the measurement modifies the  
path-integral giving privilege to the paths close to the output $a(t)$.
The propagator of a system in which the position is measured includes 
the influence of the measurement through a weight functional $w_{[a]}[x]$ 
\begin{equation}
K_{[a]}(x^{\prime\prime},\tau;x^{\prime},0)=
\int d[x] \exp \biggl\{ {i \over {\hbar}}\int_0^{\tau}L(x,\dot{x},t)dt \biggr\}
 w_{[a]}[x].
\label{PROP}
\end{equation}
The quantity $K_{[a]}(x^{\prime\prime},\tau;x^{\prime},0)$, called measurement
amplitude hereafter, can be interpreted in two alternative ways. 
If the measurement output $a$ is known, this is a transition 
amplitude from the point $x^{\prime}$ at time $t=0$ to 
the point $x^{\prime\prime}$  at time $t=\tau$ for the 
system undergoing the measurement with output $a(t)$. 
On the other side, if $x^{\prime}, x^{\prime\prime}$ are 
known, the same expression can be understood as an amplitude 
for the measurement to give the output $a(t)$ 
with the above boundary conditions.
If the system is initially in a pure state described by the wavefunction
$\psi(x,0)$, according to the first interpretation of $K_{[a]}$, 
the quantity
\begin{equation}
P_{[a]}={{\vert <\psi_{[a]}(\tau)|\psi_{[a]}(\tau)>\vert}^2 
\over {\int {\vert <\psi_{[a]}(\tau)|\psi_{[a]}(\tau)>\vert}^2 d[a]}}
\label{P}
\end{equation}
where 
\begin{equation}
\psi_{[a]}(x^{\prime\prime},\tau)={\int} K_{[a]}(x^{\prime\prime},\tau;
x^{\prime},0)\psi(x^{\prime},0)dx^{\prime}
\label{PSITAU}
\end{equation}
can be interpreted as a probability functional for the measurement output.
Due to the influence of the measurement an effective position 
uncertainty arises
\begin{equation}
\Delta a_{eff}^2= 2
{\int \tau^{-1} \int_0^{\tau} [a(t)-\bar{a}(t)]^2 dt~ 
P_{[a]}d[a] \over
\int P_{[a]}d[a] }. 
\label{DAEFFTH}
\end{equation}
\noindent
where $\bar{a}(t)$ is the most probable path which 
makes $P_{[a]}$ extremal.
The effective uncertainty $\Delta a_{eff}$ 
is greater than the instrumental error $\Delta a$
unless the system is monitored in a classical regime,
{\it i.e.} when $\Delta a \gg \sigma$ where $\sigma$ is the width of 
the initial wavefunction $\psi(x,0)$,  or in a QND way \cite{GA}.

For simplicity we represent an actual measurement with instrumental 
error $\Delta a$ lasting a time $\tau$ through a weight 
functional $w_{[a]}[x]$ 
\begin{equation}
w_{[a]}[x]=\exp \biggl \{-{1\over 2\Delta a^2 \tau}
\int_0^{\tau} [x(t)-a(t)]^2 dt \biggl \}
\label{WA}
\end{equation}
As shown in \cite{MEOP},  
the evaluation of the path-integral can be overcome by writing an effective 
Schr\"odinger equation which takes into account the influence of the 
measurement. This equation can be solved analytically in
the case of the harmonic oscillator or numerically for a generic system.
In the former situation 
the effective Lagrangian corresponds to a forced linear oscillator 
\begin{equation}
L_{eff}={m \over 2}{\dot{x}}^2-{{m\omega^2_{r}} \over {2}}x^2
-{{i\hbar} \over {\tau\Delta a^2}}a(t)x+{{i\hbar} \over 
{2\tau\Delta a^2}}a(t)^2
\label{LHO}
\end{equation}
with renormalized complex frequency 
\begin{equation}
\omega_{r}^2=\omega^2-{{i\hbar} \over {m\tau\Delta a^2}}
\label{W}
\end{equation}
Since we are interested to a finite but small value of $\tau$, 
we choose to approximate the measurement results 
with constant values $a(t)=\epsilon$ which are
the set of all the arbitrary measurement outputs in the limit 
$\tau \rightarrow 0$. 
The probability functional of the measurement path $P_{[a]}$ 
is then reduced to a function of the amplitude $\epsilon$. 
When the initial state is chosen to be Gaussian of width $\sigma$
\begin{equation}
\psi(x,0)={{ \biggl ({1\over \pi\sigma^2} \biggr )^{1/4}}
\exp\biggl (-{x^2\over 2 \sigma^2}\biggr )}
\label{PSI0}
\end{equation}
the probability $P(\epsilon)$ is also a Gaussian function 
\begin{equation}
P(\epsilon)={{1} \over {\sqrt{\pi}\Delta a_{eff}}}
\exp\biggl ( {-{{\epsilon^2} \over {\Delta a_{eff}^2}}}\biggr )
\label{PE}
\end{equation}
with an effective uncertainty
\begin{eqnarray}
\Delta a_{eff}^{-2} (\tau) = & & {1\over \Delta a^2} 
 \Re e  \biggl [1+
{\sigma^2 \over \Delta a^2 } \biggl (i{2\beta+1 \over \alpha\omega_r\tau}-
\beta^2\gamma \biggr ) \biggr ] 
\nonumber\\
& & -{\sigma^2\over \Delta a^4}
\biggl \{ \Re e \biggl [ \beta(1-{i\alpha\gamma\over \sin(\omega_r\tau)})
\biggr ] \biggr \}^2
\nonumber\\
& & \times \biggl \{ \Re e \biggr [ {1+i\alpha \tan(\omega_r\tau) \over 
1+{i\over\alpha}\tan(\omega_r\tau)} \biggr ] \biggr \}^{-1} 
\label{DAEFF}
\end{eqnarray}
having introduced $\alpha=m\omega_r \sigma^2/\hbar$, 
$\beta=[\cos (\omega_r\tau)-1]/ [\omega_r\tau \sin (\omega_r\tau)]$ and 
$\gamma=1/[1-i\alpha \cot(\omega_r\tau)]$.
Under the influence of the measurement the initial state collapses 
into a state localized around the measurement result.
If, for simplicity, we suppose that this measurement result is 
the most probable compatible with (\ref{PSI0}), {\it i.e.} 
$a(t)=0$, 
the initial Gaussian state just changes its width to
\begin{equation}
\sigma(\tau)  =  \sigma
\biggl \{ \Re e  \biggl [ {\alpha^2 \sin(\omega_r \tau)-i\alpha 
\cos(\omega_r \tau)
\over \sin(\omega_r \tau)-i \alpha \cos(\omega_r \tau)}
\biggr ] \biggr \}^{-{1 \over 2}}.
\label{SIGMA}
\end{equation}
After the measurement the state evolves according to the dynamical
law of the free, {\it i.e.} unmeasured system.
For the harmonic oscillator the state remains a Gaussian having a 
width oscillating in time 
\begin{equation}
\sigma(t+\tau)= \sigma(\tau)
\sqrt{1+\biggl [{\hbar\over m\omega\sigma(\tau)^2}\biggr ]^2 
\tan^2 (\omega t)\over{1+\tan^2 (\omega t)}}.
\label{SIGMAT}
\end{equation}
Equations (\ref{DAEFF},\ref{SIGMAT}) allow to study quantitatively 
a measurement strategy which consists of a sequence of measurements  
of duration $\tau$ equally spaced by a quiescent time $\Delta T$ 
in which no measurement is performed. 
The repeated collapses of the wavefunction during the 
measurements determine an asymptotic effective uncertainty. 
This is evident in Fig. 1 where we 
show the dependence of the effective uncertainty 
upon the number of measurements. 
After few measurements the effective uncertainty reaches an 
asymptotic value which does not depend on 
the initial state of the system.

The measurement strategy we have described can be optimized by
choosing the duration $\tau$ of each measurement and the quiescent time 
$\Delta T$ between two consecutive measurements. 
As we show in Fig. 2 the asymptotic $\Delta a_{eff}$ has minima
when $\Delta T$ is a multiple of half period of 
the harmonic oscillator $T\equiv 2 \pi / \omega$, 
{\it i.e.} in coincidence with the minima of Eq. (\ref{SIGMAT}).
The minima of $\Delta a_{eff}$ reach the instrumental 
error $\Delta a$ if $\tau \ll \tau_c$ where $\tau_c$ is the critical value 
\begin{equation}
{1 \over \tau_c}={\hbar \over m} \biggl ({1 \over \Delta a^2}+
{1 \over \sigma^2} \biggr ).
\label{TAUC}
\end{equation}
Indeed for this impulsive regime the effective uncertainty 
and the width of the collapsed wavefunction are simply written as
\begin{equation}
\lim_{\tau \to 0} \Delta a_{eff}(\tau)=\sqrt{\Delta a^2+\sigma^2}
\label{DAEFF0}
\end{equation}
\begin{equation}
\lim_{\tau \to 0}\sigma(\tau)=
\sqrt{{\sigma^2\Delta a^2}\over{\sigma^2+\Delta a^2}}.
\label{SIGMA0}
\end{equation}
In the limit of an infinite number of measurements 
the wavefunction asymptotically collapses 
to a $\delta$-function, $\Delta a_{eff}$ approaches $\Delta a$
and an ideal QND stroboscopic strategy is obtained. 
It is worth to observe that only for $\tau \sim \tau_c$ the optimal effective 
uncertainty significantly departs from $\Delta a$ while for 
$\tau \ll \tau_c$ the ideal situation $\Delta a_{eff}=\Delta a$
is very well approximated. In other words $\tau_c$ is the timescale 
which defines a quasistroboscopic behaviour of the measurement.

In Figure 1 we also compare the analytical results of Eqs.
(\ref{DAEFF},\ref{SIGMAT}) (solid curves) with the numerical
integration of the effective Schr\"odinger equation (dots). 
This allows to check the
accuracy of a numerical method (the error is less 
than 0.1\%) we use to study nonlinear systems.
We focus our attention on a system described by the Lagrangian
\begin{equation}
L={m \over 2}{\dot{x}}^2-{{m\omega^2} \over {2}}x^2-
{{\lambda} \over {4}}x^4
\label{METAL}
\end{equation}
\noindent
Also in this case the measurement strategy 
discussed for a harmonic oscillator gives rise to an asymptotic 
effective uncertainty. 
As shown in Fig. 3 the asymptotic $\Delta a_{eff}$ does not depend 
on the initial state but is a function of the measurement 
and quiescent times. 
Fig. 4 shows that in the impulsive regime $\tau \ll \tau_c$ 
the asymptotic $\Delta a_{eff}$ is an approximatively periodic function 
of the quiescent time $\Delta T$. 
The nature of these oscillations is understood in 
terms of the energy eigenvalues $E_i$ of the nonlinear oscillator. 
Indeed these eigenvalues dictate the time evolution
of the wavefunction during the quiescent intervals according to 
characteristic periods $T_{ij}/T=\hbar \omega/|E_i-E_j|$. 
Since after each measurement the wavefunction collapses around 
the measurement result, again chosen as $a(t)=\bar{a}(t)=0$, 
the relevant characteristic 
periods are those corresponding to the smallest even eigenstates. 
A WKB evaluation of the first two relevant terms 
gives $T_{20}/T=0.225$ and $T_{40}/T=0.098$. 
The fundamental time $T_{20}$ corresponds to the principal minima  
shown in Fig. 4 and $T_{40}$ corresponds to the other secondary minima.
When the quiescent time $\Delta T$ is close to a multiple of both 
$T_{20}$ and $T_{40}$ an absolute minimum is expected. This is 
what we observe in Fig. 4 at $\Delta T \simeq 3 T_{20} \simeq 7 T_{40}$.
Unlike the case of the harmonic oscillator, the general incommensurability 
of the characteristic periods $T_{ij}$ forbids to reach an optimal strategy 
with an asymptotic $\Delta a_{eff}=\Delta a$ also in the impulsive regime.

Two problems recently under investigation also from a 
phenomenological point of view may take advantage of our approach. 
Firstly, it has been suggested that the hypothesis 
of realism underlying classical mechanics can be confronted in the 
macroscopic domain with quantum predictions, namely the existence 
of macroscopic distinguishable states, measuring the magnetic flux 
in a rf-SQUID \cite{LE,LEGA,TE}.
In this proposal there is also the 
assumption of a so called non-invasive measurement whose role has been 
criticized due to a potential incompatibility with limitations 
in the accuracy of any measurement dictated by the uncertainty 
principle \cite{BA1,PE,EL}. 
Secondly, quantum Zeno effect has been proposed 
to account for an experiment involving inhibition of optical 
transitions between quantum states due to the measurement process 
\cite{ITA} but some debate in the literature followed on 
the validity of such an interpretation \cite{BA2,FE}.
A quantitative study of both these problems
is possible within the framework we propose here.

%%%%%%%%%%%%%%%%%%%%%%%%%%%%%%%%% ACKNOWLEDGMENTS
\acknowledgments 
This work has been supported by INFN, Italy.
%
%%%%%%%%%%%%%%%%%%%%%%%%%%%%%%%%% REFERENCES LIST
%

%
%%%%%%%%%%%%%%%%%%%%%%%%%%%%%%%%% FIGURE CAPTIONS
%
\begin{figure}
\centerline{\hbox{\psfig{figure=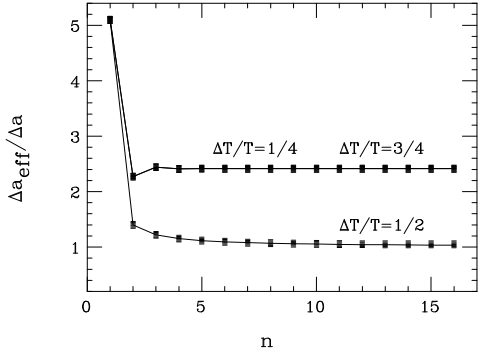,width=9.5cm,angle=90}}}
\caption{Effective uncertainty $\Delta a_{eff}$ versus the number
$n$ of repeated measurements in the case of a harmonic oscillator.
Three different quiescent times $\Delta T$ are shown:
circles are numerical results and solid lines are the analytical 
result of Eqs.\ (\protect\ref{DAEFF} - \protect\ref{SIGMAT}). 
Note that the cases $\Delta T/T=1/4$ and $\Delta T/T=3/4$ coincide.
We put $2m=\hbar=\omega=1$, $\Delta a=1$, $\sigma=5$ and $\tau/T=10^{-5}$.
\label{FIG1}}
\end{figure}
  
\begin{figure}
\centerline{\hbox{\psfig{figure=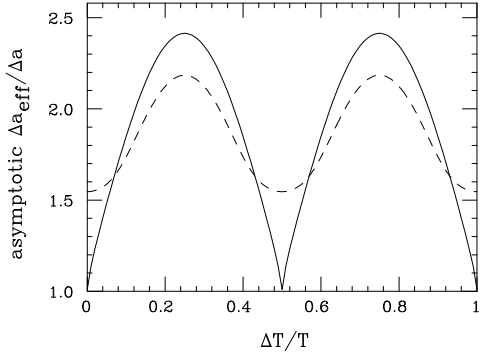,width=9.5cm,angle=90}}}
\caption{Dependence of the asymptotic effective uncertainty $\Delta a_{eff}$
on the quiescent time $\Delta T$ for the harmonic oscillator. The different 
curves are relative to different measurement times $\tau$: two solid 
coincident lines are for $\tau=0$ and $\tau=10^{-5} T$, the dashed 
line is for $\tau=10^{-1} T \sim \tau_c$. 
\label{FIG2}}
\end{figure}

\begin{figure}
\centerline{\hbox{\psfig{figure=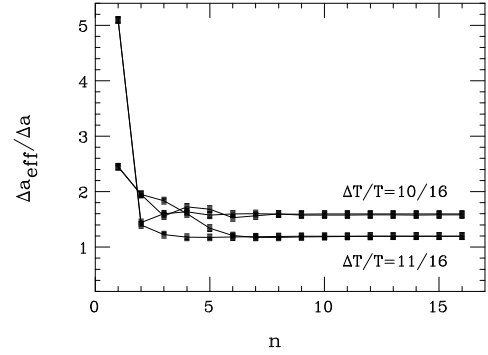,width=9.5cm,angle=90}}}
\caption{Effective uncertainty $\Delta a_{eff}$ versus the number $n$ 
of repeated impulsive measurements for the anharmonic oscillator
 with $\lambda=4$. 
Two different quiescent times $\Delta T$ are shown.
circles correspond to an initial Gaussian state with $\sigma=5$ and 
crosses are relative to a double peaked initial state. The solid 
lines are an eye guide. We put $\tau/T=10^{-5}$.
\label{FIG3}}
\end{figure}

\begin{figure}
\centerline{\hbox{\psfig{figure=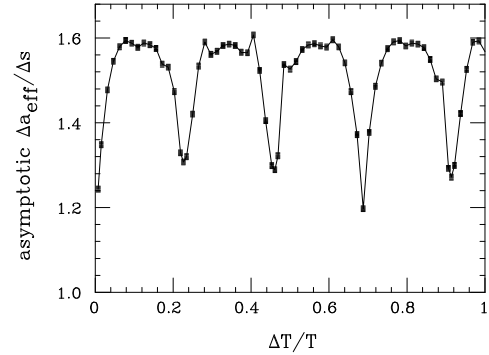,width=9.5cm,angle=90}}}
\caption{Dependence of the asymptotic effective uncertainty $\Delta a_{eff}$ 
on the quiescent time $\Delta T$ for the anharmonic oscillator 
with $\lambda=4$.  
\label{FIG4}}
\end{figure}

\end{document}